# Electron-phonon interaction effects on the direct gap transitions of nanoscale Si films

V.K. Kamineni[*] and A.C. Diebold

College of Nanoscale Science and Engineering, University at Albany, 255 Fuller Road, Albany NY 12203, USA

ABSTRACT. This study shows that the dielectric function of crystalline Si quantum wells ($c$-Si QW) is influenced by both carrier confinement and electron-phonon interactions. The energy shifts and lifetime broadening of the excitonic $E_1$ direct gap transition of $c$-Si QWs from 2 to 10 nm are found to have a significant dimensional and temperature dependence that can be traced to changes in the phonon dispersion of nanoscale films. The influence of electron-phonon interactions on the dielectric function was verified by altering the phonon dispersion using different dielectric layers above a 5 nm $c$-Si QW.

MANUSCRIPT TEXT

The linear optical response (dielectric function) of nanoscale crystalline semiconductor materials is often understood in terms of the effects of quantum confinement (QC).[1-3] For instance, the experimentally measured blue shift in the direct gap absorption or critical point (CP) of indirect band gap semiconductors with increased dimensional confinement have been related to quantum confinement of electrons.[3,4] Most of this previous work relates the blue shift in the $E_1$ CP of silicon to QC of carriers, and disregard the role of electron-phonon interactions. Here, we provide experimental evidence of the effect of change in phonon dispersion on the dielectric function of quantum wells (QWs) of single crystal silicon (*c*-Si) films on insulator structures. $HfO_2$ covered c-Si QWs show redshifts in the $E_1$ CP energy and thus quantum confinement does not play a dominant role. This paper presents a temperature dependent study of the dielectric function of *c*-Si QWs, which shows that electron-phonon interactions also play a key role in the optical properties of semiconductor nanostructures. We further demonstrate that the dielectric function will change with different phonon dispersions using *c*-Si QWs with three different surface layers: native oxide, thick silicon dioxide, and hafnium oxide.

In order to describe the changes in optical response observed for nanoscale *c*-Si films, a brief discussion of the dielectric function is useful. The optical response of Si is dominated by two direct band transitions where there is a constant energy difference between the valence band (VB) and the conduction band (CB). This high joint density of states results in sharp features in the optical absorption known as critical points (e.g., $E_1$ and $E_2$ CP). The $E_1$ CP is due to the transition from $\Lambda_3$ VB to $\Lambda_1$ CB between the wave vectors k = $(2\pi/a_o)$ <¼,¼,¼> and k = $(2\pi/a_o)$ <½,½,½>.[5] *Ab initio* calculations of the dielectric function of bulk Si at zero Kelvin show that the oscillator strength and transition energy (~3.4 eV) of the $E_1$ CP are strongly impacted by electron-hole (excitonic) interactions.[6,7] Cardona et al. found that the temperature dependence of the indirect band gap and the direct transitions in bulk semiconductors is due to both thermal expansion and electron-phonon interactions.[5,8,9] Here, we show that the shifts in the $E_1$ CP energy and lifetime broadening of nanofilms of Si are strongly influenced by electron-phonon interactions. The $E_0'$ transition (degenerate CP of the $E_1$ CP that can be resolved[10]) at the $\Gamma$-point also shows an energy shift. The $E_2$ and other CPs do not have a significant energy shift due to change in thickness, but do show changes in oscillator (absorption intensity) strength. Another aspect of decreasing dimensions is the change in CB and VB energies. We note that the experimental studies of the indirect band gap indicate that significant increase in gap between the CB and VBs occur at and below 3 nm.[11] The films used in this study are between 10 nm and 2 nm in thickness.

To study the impact of electron-phonon interactions, $c$-Si QWs were fabricated from silicon-on-insulator (SOI) wafers. A series of 300 mm SOI wafers (700Å $c$-Si/1400Å oxide/bulk Si) underwent a wet oxidation step and etching in dilute HF (100:1) to remove the bulk of the top Si thickness. This was followed by cyclic steps of dry oxidation and etching in dilute HF (300:1) to achieve the desirable thickness (~ 2 to 10 nm) of $c$-Si QWs, commonly referred to as extremely thin SOI (ETSOI). After each oxidation and etching step the thickness was monitored using an in-line spectroscopic ellipsometry (SE). It was previously demonstrated that crystal quality ($c$-Si) is maintained in the QWs after thinning as verified by high-resolution transmission electron microscopy.[4] Raman spectroscopy and high-resolution x-ray diffraction demonstrated that the $c$-Si QWs had no measurable strain.[2] To study the change in optical response of $c$-Si QWs due to the top dielectric layer the following sets of wafers were fabricated: (i) a set with native oxide (~1nm), (ii) a set with thicker thermal oxide, and (iii) a set with hafnium oxide (~10 nm). SE was used to determine the dielectric function. Temperature dependent studies of the films with the native oxide were used to demonstrate the impact of changes in phonon dispersion on the $E_1$ CP energy and broadening. The low-temperature SE measurements were performed on dual rotating compensator ellipsometer (J.A.Woollam) equipped with a variable temperature cryostat, with the temperature ranging from 4 K to 300 K. Direct space analysis was used to extract the energy and lifetime broadening, by fitting the second derivative of the dielectric function of each parabolic CP with a Lorentzian line shape shown in Eq. 1. The second derivative of the dielectric function was taken to enhance the CP structure and suppress the baseline effects.[12]

$$\frac{d^2\varepsilon}{dE^2} = \begin{cases} \mu(\mu+1)Ae^{i\beta}\left(E-E_g+i\Gamma\right)^{-\mu-2}, \mu \neq 0 \\ Ae^{i\beta}\left(E-E_g+i\Gamma\right)^{-2}, \mu = 0 \end{cases} \quad (1)$$

Where $A$ is the amplitude, $\beta$ is the phase angle, $E_g$ is the threshold energy, $\Gamma$ is the broadening and $\mu$ is the order of singularity. The value of $\mu$ is based on the type of CP and it is ½, 0 and -½ for one-dimensional (1D), two-dimensional (2D) and three-dimensional (3D) one-electron transitions, respectively.[4] While in the case of $E_1$ CP, its value is 1 due to the discrete excitonic nature of the CP.

The first evidence that QC is not always a dominant effect determining the energy of direct transitions is the small red shift versus decreasing thickness of the $E_1$ CP (second derivative in Fig. 1) of hafnium oxide covered *c*-Si QWs as shown in the inset Fig. 1 (a). This red shift is unexpected, and indicates that the electrons and holes excited at the $E_1$ CP of the hafnium oxide covered *c*-Si QWs are more strongly influenced by electron-phonon interactions than quantum confinement of carriers. The lifetime broadening shown in the inset Fig. 1 (b) of the $E_1$

CP increases for thinner $c$-Si, due to change in the electron-phonon interactions induced by changes in the phonon modes of thin films.

Next the effect of electron-phonon interactions on the $E_1$ CP of $c$-Si QWs with native oxide is studied. The thickness and temperature dependence of the imaginary part of dielectric function of $c$-Si films with a native oxide surface at room temperature for 9 nm, 7 nm, and 2 nm $c$-Si films is shown in Fig. 2(a)-2(c) between 2 eV and 5 eV as measured by SE. The changes in optical response with temperature are a strong function of film thickness. The oscillator strength of the dielectric function of bulk and 9 nm silicon is stronger at the $E_2$ CP than the $E_1$ CP at all temperatures. This is clearly not the case for the 7 nm or less $c$-Si QW films with the native oxide surface layer. As temperature decreases, the oscillator strength of the $E_1$ CP of the 7 nm $c$-Si peak increases above that of the $E_2$ CP.

The temperature dependence of the $E_1$ CP energy and its lifetime broadening are shown in Figs. 3(a) and 3(b) respectively. The temperature dependence of the energy and lifetime broadening of the $E_1$ CP for all the thicknesses of $c$-Si QWs follow a similar trend with temperature as observed for bulk Si. The broadening of the $E_1$ CP for 9 nm, 8 nm, and 7 nm $c$-Si films is relatively small, while the 2.5 and 2 nm $c$-Si films also have strong absorption at the $E_1$ CP; with a lifetime broadening that is more pronounced. There are two main contributions to the temperature dependence of the band gap and direct transitions.[5,8,9,12] One is the temperature dependence of the lattice spacing, which can be directly related to the change in band structure due to thermal expansion coefficient. The other term is a result of electron-phonon interactions. The electron-phonon interactions are due to contributions from the Debye-Waller term and Fan or "self-energy" term. Cardona used a temperature dependent electron-phonon spectral function along the $\Lambda$ direction for initial states to show the difference in how optical and acoustic phonon interactions impact CB and VB states.[5] These theoretical calculations of the electron-phonon interaction terms for bulk Si indicate that both optical and acoustic phonon contribute to the temperature dependence of the broadening, while acoustic phonon more strongly impact the temperature dependence of the $E_1$ CP energy. Furthermore, these theoretical calculations for bulk Si indicate that electron-phonon interactions during optical transitions differ between the VB and CB. Our goal is to experimentally investigate the impact of electron-phonon interactions on *c*-Si QWs. The electron-phonon interactions arise due to the particle nature (boson) of phonons where the emission and absorption of phonons by electrons occurs. The probability of phonons occupying each phonon mode is temperature dependent and the distribution follows Bose-Einstein statistics. The Bose-Einstein distribution predicts that the average number of phonons increases with increase in temperature and thereby increasing the electron-phonon interactions. Thus the energy shift due to both thermal expansion and electron-phonon interactions can be fit to a theoretically inspired phenomenological form based on the Bose-Einstein statistical factor[12] for phonons with an average interaction frequency '$K_B\theta/\hbar$' in Eq. 2.

$$\left(\frac{\partial E_g}{\partial T}\right)_p = E_B - a_B\left(1 + \frac{2}{e^{\theta/T}-1}\right) \qquad (2)$$

It has been reported by Lautenschlager et al. that the lower the average phonon frequency,[5] the stronger the contribution of acoustic phonons on the $E_1$ CP. Table I shows that the contributions of acoustic phonons increases with decreasing *c*-Si QW thickness. This temperature dependency is a strong evidence of the importance of electron-phonon interactions in determining the optical properties of nanoscale *c*-Si. This observation supersedes previous suppositions that quantum confinement of carriers played a dominant role in the blue shifting of the $E_1$ CP with decreasing thickness.[2]

The increasing contribution of acoustic modes to the temperature dependence of the E1 CP with decreasing QW thickness can be understood using recent theoretical modeling of the lattice dynamics of silicon nanofilms. Nanofilms have a number of additional optical phonon branches.[13] Modeling based on the adiabatic bond charge method has shown that the lowest lying optical phonon (LOP) mode follows $\omega(\text{cm}^{-1}) = 188.25/\xi^{3/4}$ where $\xi = d/a_0$ at the zone center. Where d is the film thickness and $a_0$ the lattice constant 0.543 nm.[13] The LOP frequency increases by a factor of two between $\xi =10$ and 4, and the Bose-Einstein occupation probability for this phonon decreases by more than a factor of two. Thus, the increasing influence of the acoustic modes on the $E_1$ CP. Although the phonon dispersion of nanofilms are no longer flat for $\xi>2$, optical phonons across all wave vectors should play a larger role for thicker films.

The effect of changing the energy of the phonon modes in *c*-Si QW was studies using top layers of native oxide (~ 1nm), thicker thermal oxide, and hafnium oxide (~10 nm). The challenges in fabrication result in slight differences (± 2 Å) between wafers in the thickness of the nanoscale *c*-Si film. In Fig. 4 we show a change in the room temperature dielectric function of 5 nm *c*-Si each with the three different surface layers. Although the depth of top QW barrier is less for the hafnium oxide covered *c*-Si QW (2.7 eV) when compared to the $SiO_2$ covered *c*-Si QW (5.7 eV), all three samples have nearly identical thickness and thus should have the same amount of carrier confinement at room temperature. We interpret the change in the dielectric function representative of the effect of changes in the acoustic phonon modes in the three samples. The acoustic phonon energies of the *c*-Si QW change with the stiffness of the surface layers and the Young's modulus of $HfO_2$ (~370 GPa) is greater than that of $SiO_2$ (~75 GPa). It provides further proof of the importance of electron-phonon interactions on the optical response of nanoscale single crystal films, and points to the strong possibility that the same will be true for nanowires and nanodots. Aspnes et. al have shown that changes in the surface of bulk Si can impart small changes in the dielectric function especially to the $E_1$ CP.[14] The lifetime of the final state has been used to explain the ~10 meV difference in measured CPs between (111) and (001) orientations. We note that the crystal orientation dependent changes in the energy of the $E_1$ CP are smaller than those observed here.

Understanding the physics behind dimensional changes in the dielectric function has both fundamental and practical implications. For example, one can design structures that produce desired optical properties or provide accurate measurements of material dimensions. Measurement of nanoscale *c*-Si film thickness by SE requires knowledge of the thickness dependence of its dielectric function. We have presented experimental evidence that the optical properties of single crystal semiconductor nanoscale structures are altered by both electron (hole) confinement and electron-phonon interactions. This study shows that the electron-phonon interactions play a key role in the optical response of nanoscale, single crystal films of Si. Temperature dependent changes in the energy and broadening of the $E_1$ CP are consistent with interpretation of changes in the $E_1$ CP of bulk Si in terms of electron-phonon interactions. Based on the theoretical studies of Cardona and co-workers for bulk Si, we also attribute changes in the energy and lifetime broadening of the $E_1$ CP of nanoscale Si to electron-phonon interactions. It is noted that although the energy shift was strongly influenced by acoustic phonon, optical phonons also play a role in the energy shift with temperature, and play a significant role in changes in broadening. Here we provide experimental evidence of the importance of electron-phonon interactions on optical properties by altering the phonon dispersion of *c*-Si QW films by the addition of surface layers of $SiO_2$ and $HfO_2$. This research suggests that the optical response of *c*-Si QWs can be selectively altered through fabrication of film stacks with dielectric layers having appropriate elastic properties. This opens new avenues into the fabrication of nanoscale structures[15] with new optical properties. Also, the mobility of *c*-Si QWs is impacted by electron-phonon scattering, and appropriate dielectric films surrounding the *c*-Si QWs could be used to improve carrier transport in devices.

ACKNOWLEDGMENTS

We gratefully acknowledge discussions with Brennan Peterson as well as the encouragement of Manuel Cardona concerning the interest in the optical properties of nanoscale semiconductor structures. We acknowledge the assistance of Tianhao Zhang and Eric Bersch, and the sample fabrication engineering by Steve Gausephol and Hui-feng Li. We also acknowledge support from the New York Center for National Competitiveness in Nanoscale Characterization (NC$^3$) and Center for Nanoscale Metrology.

| Thickness (nm) | θ (K) |
|---|---|
| 9.2 | 406 (167) |
| 8 | 356 (38) |
| 7 | 231 (58) |
| 2.5 | 98.5 (46) |
| 2 | 90 (65) |

TABLE I: Average phonon frequency (90 % confidence limits) at various *c*-Si film thicknesses

FIGURE CAPTIONS

**Figure 1.** Second derivative of the imaginary part of the dielectric function of *c*-Si QW below a HfO$_2$ (~9 nm) surface layer at 300 K. Inset: (a) Energy and (b) lifetime broadening ($\Gamma$) of the E$_1$ CP extracted using direct space analysis.

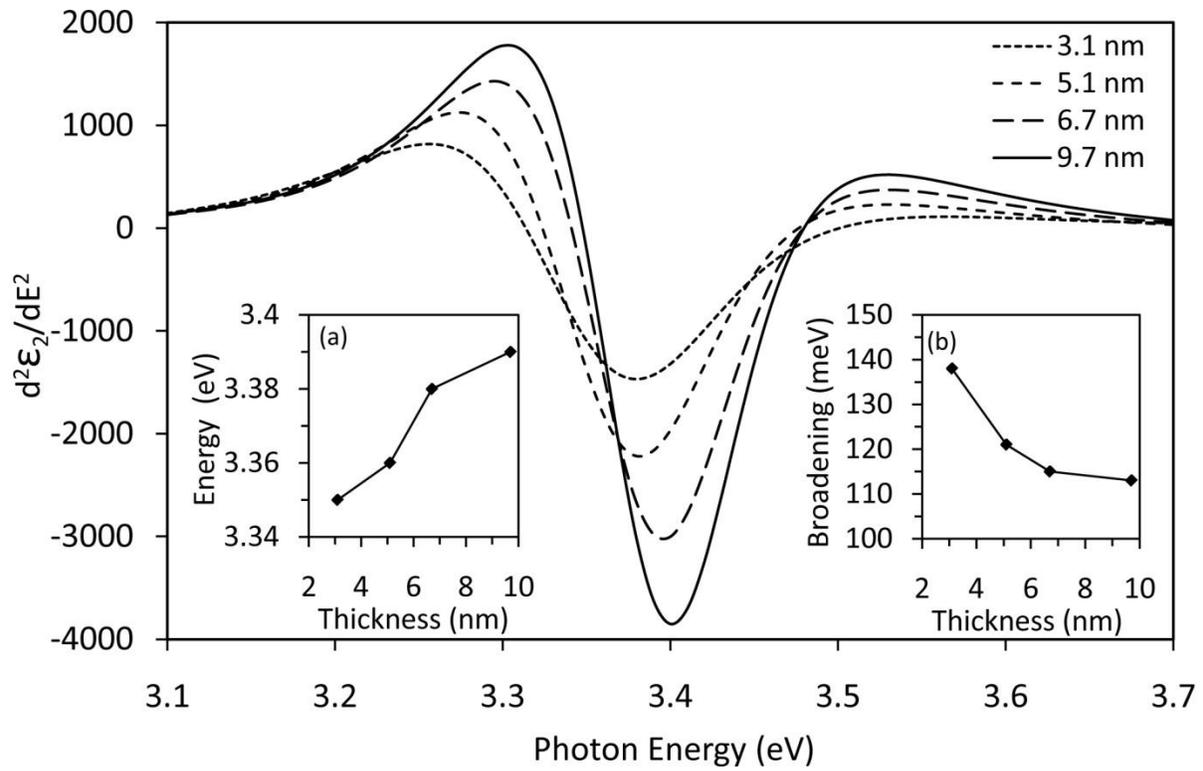

**Figure 2.** Imaginary part of the dielectric function ($\varepsilon_2$) for (a) 9 nm, (b) 7 nm and (c) 2 nm of *c*-Si QW at a series of temperatures.

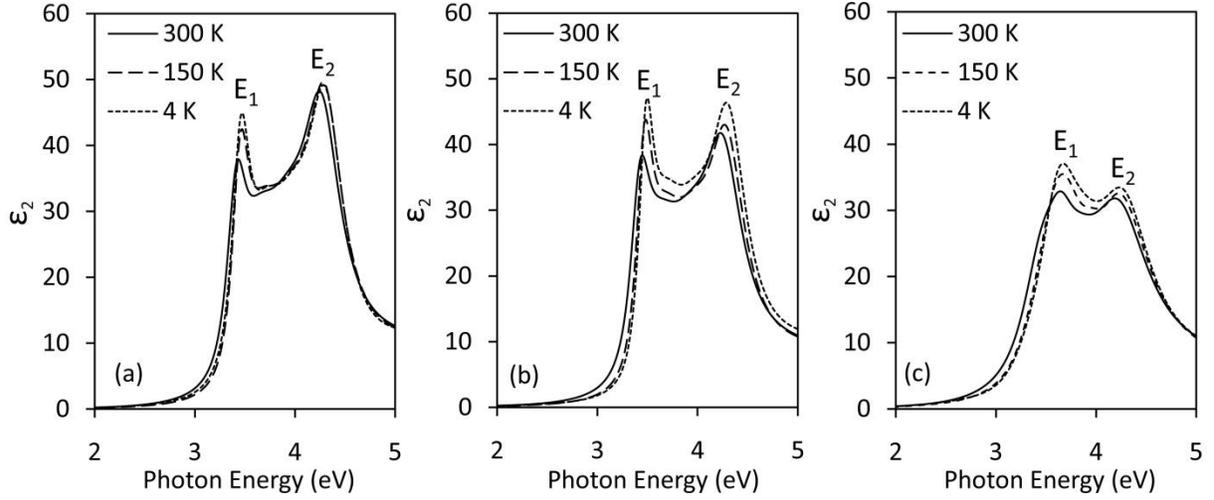

**Figure 3.** (Color online). The thickness and temperature dependence of the (a) energy and (b) lifetime broadening of $E_1$ CP (95 % confidence limits for the energy and broadening are less than 10 meV).

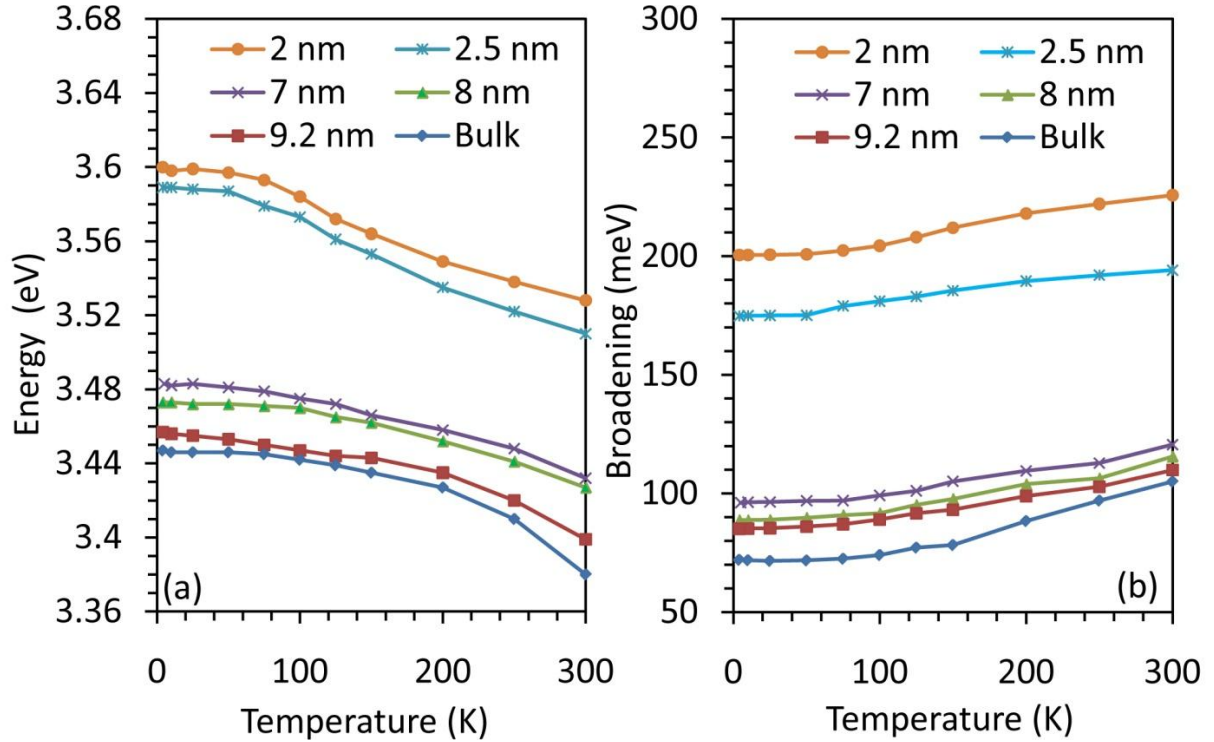

**Figure 4.** Imaginary part of the dielectric function of *c*-Si QWs (~ 5 nm) with native oxide, 20 nm $SiO_2$, and 10 nm $HfO_2$ with a $SiO_2$ interfacial layer. Note the clear shift in the energy of the $E_1$ CP and the changes in lifetime broadening.

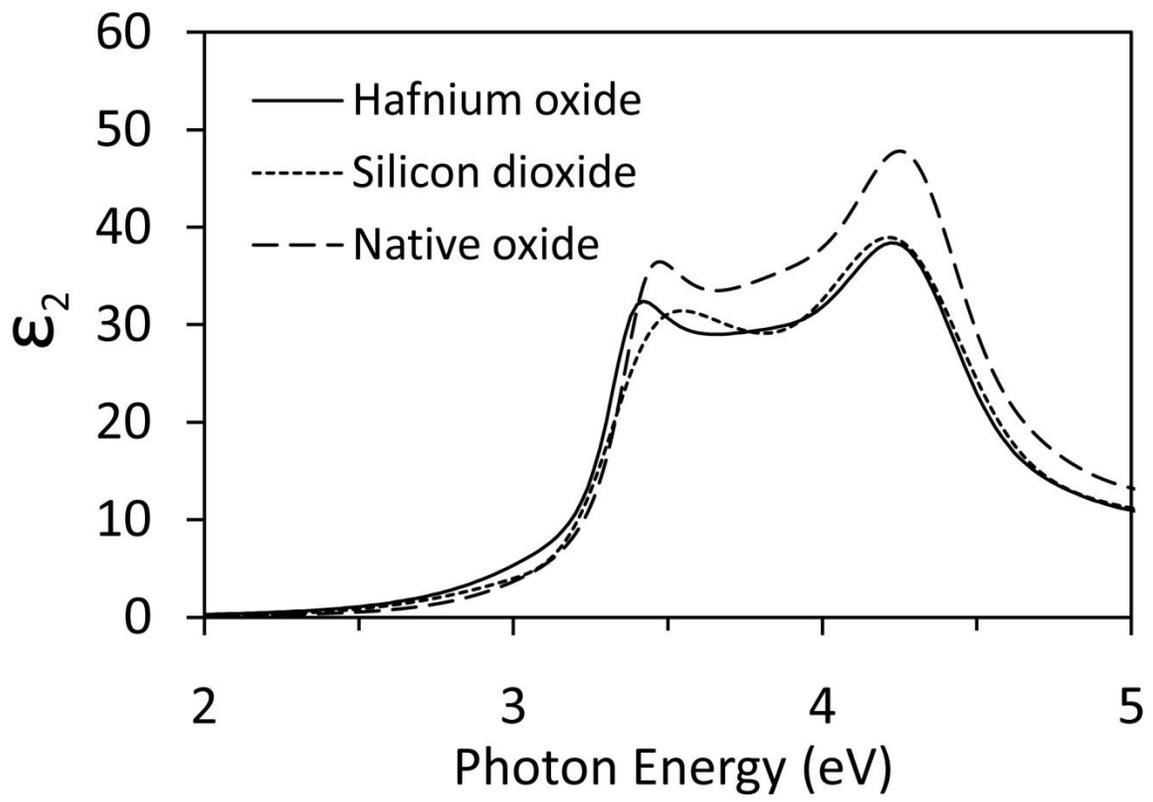